\documentclass{article}

\usepackage{arxiv}
\usepackage{cite}
\usepackage[utf8]{inputenc} 
\usepackage[T1]{fontenc}    
\usepackage{hyperref}       
\usepackage{url}            
\usepackage{booktabs}       
\usepackage{amsfonts}       
\usepackage{nicefrac}       
\usepackage{microtype}      
\usepackage{graphicx}
\usepackage[sort,compress,square,numbers]{natbib}
\usepackage{doi}
\usepackage[nottoc]{tocbibind}
\usepackage{float}
\usepackage{siunitx}
\usepackage{soul}
\usepackage{xcolor}
\usepackage[normalem]{ulem}

\title{An Efficient Compact Blazed Grating Antenna for \\Optical Phased Arrays}

\author{ \href{https://orcid.org/0000-0001-7730-3489}{\includegraphics[scale=0.06]{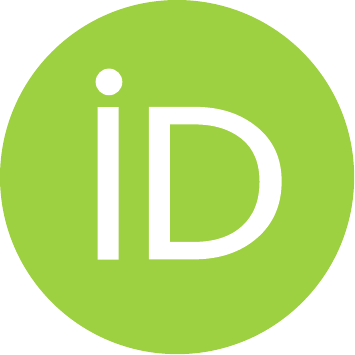}\hspace{1mm}Henna Farheen}\\
	Paderborn University\\
	Theoretical Electrical Engineering\\
	Warburger Str. 100, 33098 Paderborn, Germany \\
	\texttt{henna18@campus.uni-paderborn.de} \\
	\And
    \hspace{1mm}Suraj Joshi \\
	Paderborn University\\
	Theoretical Electrical Engineering\\
	Warburger Str. 100, 33098 Paderborn, Germany \\
	\texttt{surajj@campus.uni-paderborn.de} \\
	\And
	\href{https://orcid.org/0000-0002-5950-6618}{\includegraphics[scale=0.06]{orcid.pdf}\hspace{1mm}J. Christoph Scheytt}\\
	Paderborn University\\
	System and Circuit Technology\\
	F\"{u}rstenallee 11, 33102 Paderborn, Germany\\
	\texttt{christoph.scheytt@hni.uni-paderborn.de} \\
	\And
	\href{https://orcid.org/0000-0001-6431-746X}{\includegraphics[scale=0.06]{orcid.pdf}\hspace{1mm}Viktor Myroshnychenko} \\
    Paderborn University\\
	Theoretical Electrical Engineering\\
	Warburger Str. 100, 33098 Paderborn, Germany \\
	\texttt{viktor.myroshnychenko@uni-paderborn.de} \\
	\And
	\href{https://orcid.org/0000-0001-7059-9862}{\includegraphics[scale=0.06]{orcid.pdf}\hspace{1mm}Jens F\"orstner} \\
    Paderborn University\\
	Theoretical Electrical Engineering\\
	Warburger Str. 100, 33098 Paderborn, Germany \\
	\texttt{jens.foerstner@uni-paderborn.de} \\
}

\date{}

\renewcommand{\headeright}{}
\renewcommand{\undertitle}{}
\renewcommand{\shorttitle}{An Efficient Compact Blazed Grating Antenna for Optical Phased Arrays}

\hypersetup{
pdftitle={An Efficient Compact Blazed Grating Antenna for Optical Phased Arrays},
pdfkeywords={Dielectric antenna, phased arrays, field of view, directivity},
}

\begin{document}
\maketitle

\begin{abstract}
Phased arrays are vital in communication systems and have received significant interest in the field of optoelectronics and photonics, enabling a wide range of applications such as LiDAR, holography, wireless communication, etc. In this work, we present a blazed grating antenna that is optimized to have upward radiation efficiency as high as 80\% with a compact footprint of \SI[mode=text]{3.5}{\micro\metre}$\times$\SI[mode=text]{2}{\micro\metre} at an operational wavelength of \SI[mode=text]{1.55}{\micro\metre}. Our numerical investigations demonstrate that this antenna in a 64$\times$64 phased array configuration is capable of producing desired far-field radiation patterns. Additionally, our antenna possesses a low side lobe level of -9.7\,dB and a negligible reflection efficiency of under 1\%, making it an attractive candidate for integrated optical phased arrays.   
\end{abstract}

Integrated optical antennas in the transmitting mode are devices that couple localized power into free propagating radiation in their surrounding medium, where the freely radiating electromagnetic waves interfere in the far-field \cite{balanis2015antenna}. These antennas pave the way to numerous integrated photonic systems like optical phased arrays (OPAs) \cite{fatemi2020breaking}, wireless communication \cite{poulton2019long}, coherent imagers \cite{aflatouni2015nanophotonic}, etc. These systems are commonly realized using silicon photonic integration, which is highly compatible with the complementary metal-oxide-semiconductor (CMOS) process, thus furnishing high-yield, low-cost commercial systems \cite{chung2017monolithically}. In particular, for OPAs, this has proven to be extremely beneficial in incorporating a large number of antennas that aid in generating desired far-field radiation patterns \cite{sun2013large1}.    

Conventionally, the performance of an OPA is strongly affected by the number of elements and the spacing between them, which control the angular resolution, beamwidth, and the periodicity of the grating lobes. Many of these arrays are implemented as 1D-OPAs, as they can tightly assemble long and narrow radiators \cite{abiri20181,kossey2018end,poulton2017large}. However, such a configuration can perform beam steering solely in one direction unless used with a highly precise tunable laser \cite{fatemi2019nonuniform}. On the other hand, 2D-OPAs are capable of beam steering in two directions via phase tuning at the operational wavelength. Nonetheless, they come at the cost of a limited field of view (FOV), i.e., the grating-lobe-free region for beam steering \cite{sun2013large1}, due to the large footprint of the radiating elements. Having an inter-element spacing greater than half the optical wavelength results in undesirable grating lobes in the far-field radiation patterns. However, it has been demonstrated that the issue of limited beam steering can be partially mitigated using a sparse array configuration, integrating a large number of elements in an N$\times$N grid in a non-uniform fashion with minimal inter-element spacing, taking into account the implications of waveguide routing \cite{fatemi2019nonuniform}.  An alternative approach is to employ circularly symmetric array configurations, exploiting the properties of a zero-order Bessel-like intensity distribution in the far-field which implicates a visible region with no grating lobes \cite{khachaturian2022discretization}. 

Together, challenges like large FOV, high radiation efficiency, and reduced form factor with compact unit cells continue to create a surge for devising new radiator designs that can fulfill these requirements \cite{farheen2023increasing}. Recently, it has been shown that dielectric horn antennas possess highly directive fields \cite{farheen2022optimization,farheen2022broadband,leuteritz2021dielectric}, while blazed gratings are capable of near vertical high radiation efficiency \cite{liu2022circular,kazemian2021optimization,benedikovivc2022circular,melati2020design}. In this letter, we report on a compact blazed grating horn antenna which is optimized for a wavelength of \SI[mode=text]{1.55}{\micro\meter} to produce a high upward radiation efficiency with a broadside emission angle constraint. This is done by performing full-wave numerical simulations utilizing the finite element method (FEM) in conjunction with a hybrid optimization routine that includes particle swarm optimization followed by the trust region method. The optimization goal function, the upward efficiency ($\eta_{up}$), can be defined as
\begin{equation}
\eta_{up}=\frac{\int_{0}^{2\pi}\int_{0}^{\pi/2} P_{rad}(\theta,\varphi) d\theta d\varphi}{P_{in}},
\label{eq::eq1}
\end{equation}
where $P_{rad}$ is the power radiating in the defined computation domain, $P_{in}$ is the input optical power, $\theta$ is the polar angle, and $\varphi$ is the azimuthal angle. 

\begin{figure*}[t!]
	\centering
	\includegraphics[width=17cm]{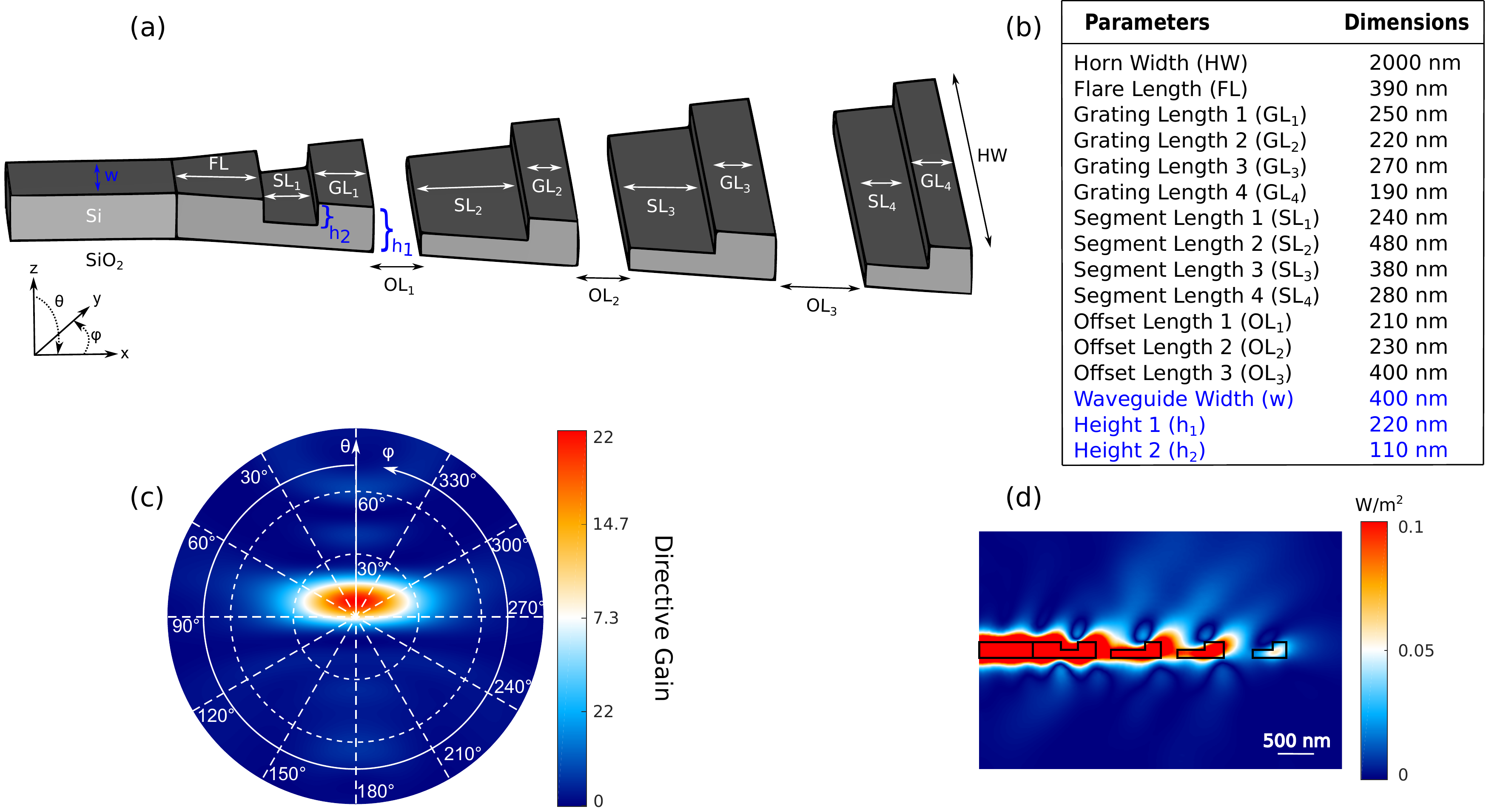}
	\caption{(a) Schematic representation of the optimized horn antenna, highlighting the parameters used in the optimization. (b) Values of the design parameters obtained for the optimized antenna. (c) Calculated angular linear directive gain distribution of the optimized antenna exhibiting a directivity of $D=22$ at $\theta=8^{\circ}$. (d) Calculated near-field distribution of the power flow for the optimized antenna in the $xz$-plane at $y=0$.}
	\label{fig:figure1}
\end{figure*}

Our proposed antenna consists of a waveguide-fed silicon horn antenna ($n_{Si}=3.4$) comprising four gratings in a silicon dioxide ($n_{SiO_2}=1.44$) environment. The first grating is a partially etched trapezoidal grating with a U-shape, while the other three gratings are L-shaped gratings. Fig.~\ref{fig:figure1}a illustrates the schematic of the antenna with an orientation along the $xy$-plane. Along the $x$-axis, which is the direction of propagation, the different lengths are classified as grating lengths (GL$_x$), offset lengths (OL$_x$), segment lengths (SL$_x$), and flare length (FL), which is the initial taper length. These parameters along with the horn width (HW), result in thirteen optimization parameters whose optimal values are shown in Fig.~\ref{fig:figure1}b. The values in blue highlight the fixed dimensions of the structure like the width of the feeding waveguide (w), height of the antenna ($h_1=220$\,nm), and height of the partial etch ($h_2=110$\,nm) in the U-shaped grating. The optimized structure has a compact footprint of \SI[mode=text]{3.5}{\micro\metre}$\times$\SI[mode=text]{2}{\micro\metre} with a linear directivity of 22, as shown in the calculated linear directive gain distribution in Fig.~\ref{fig:figure1}c, centered at $\theta=8^\circ$. The antenna demonstrates a low side lobe level of -9.7\,dB. Furthermore, Fig.~\ref{fig:figure1}d shows the near-field power distribution in the $xz$-plane at $y=0$. Along the length of the radiator, the input power constructively interferes in the upper hemisphere due to the multi-layer up-down asymmetries, while destructive interference dominates in the lower hemisphere.

\begin{figure}[t!]
	\centering
	\includegraphics[width=8cm]{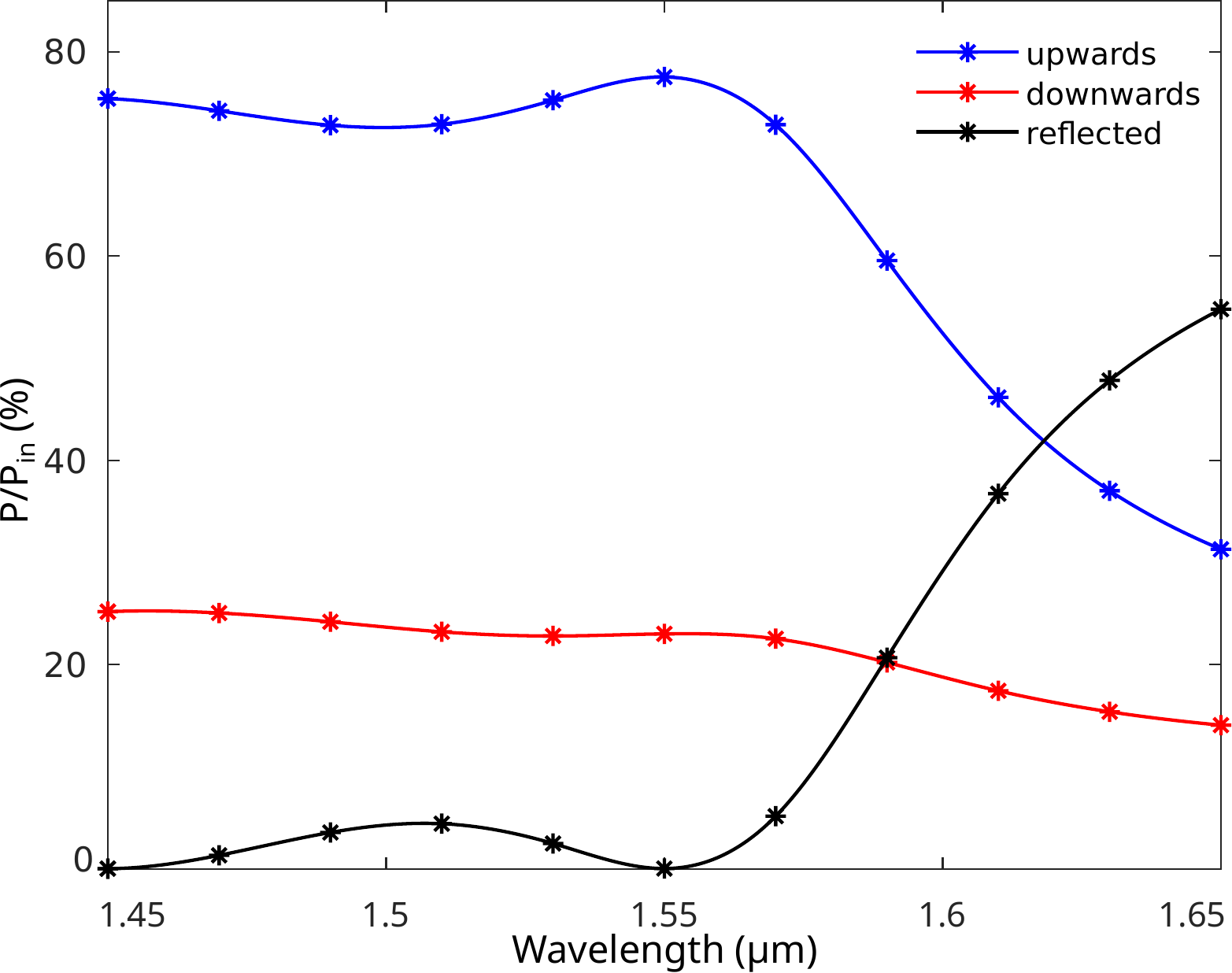}
	\caption{Calculated optical radiation efficiencies of the optimized antenna as functions of the wavelength.}
	\label{fig:figure2}
\end{figure}

To get an insight into the broadband behavior of the antenna, the radiation efficiencies emitting up and down as functions of the wavelength are shown in Fig.~\ref{fig:figure2}. The blue and red curves highlight the upward and downward radiation efficiencies, respectively. As the antenna is optimized for \SI[mode=text]{1.55}{\micro\metre}, it has its best performance at this wavelength and is sensitive to the longer wavelengths thereon. The structure maintains a low downward efficiency throughout the operating range. The upward radiation is higher for shorter wavelengths and reduces drastically for longer wavelengths, which results in an increased reflection to the waveguide, whereas maintaining a consistently low downward radiation efficiency. This implies that the antenna design works well in breaking the up-down symmetry, thus preventing more downward radiation. To lower the reflection efficiency over a wide range of wavelengths, a sub-wavelength grating (SWG) design approach can also be incorporated, as demonstrated in Ref.~\cite{benedikovivc2022circular}. At \SI[mode=text]{1.55}{\micro\metre}, the antenna exhibits a high upward radiation efficiency of approximately 80\%, partly influenced by the partial etch in the U-shaped grating which helps create a phase difference between the upward and downward propagating radiation \cite{sun2013large1}. The aperiodic L-shaped trapezoidal diffraction gratings further strengthen the up-down asymmetry along the entire length of the structure. This reinforces the constructive interference in upward direction, destructive interference in downward direction, and consequently, reduces the in-plane propagation. Furthermore, the antenna has almost no power that is reflected back into the feeding waveguide, thus making it highly efficient.

\begin{figure*}[t!]
	\centering
	\includegraphics[width=\textwidth]{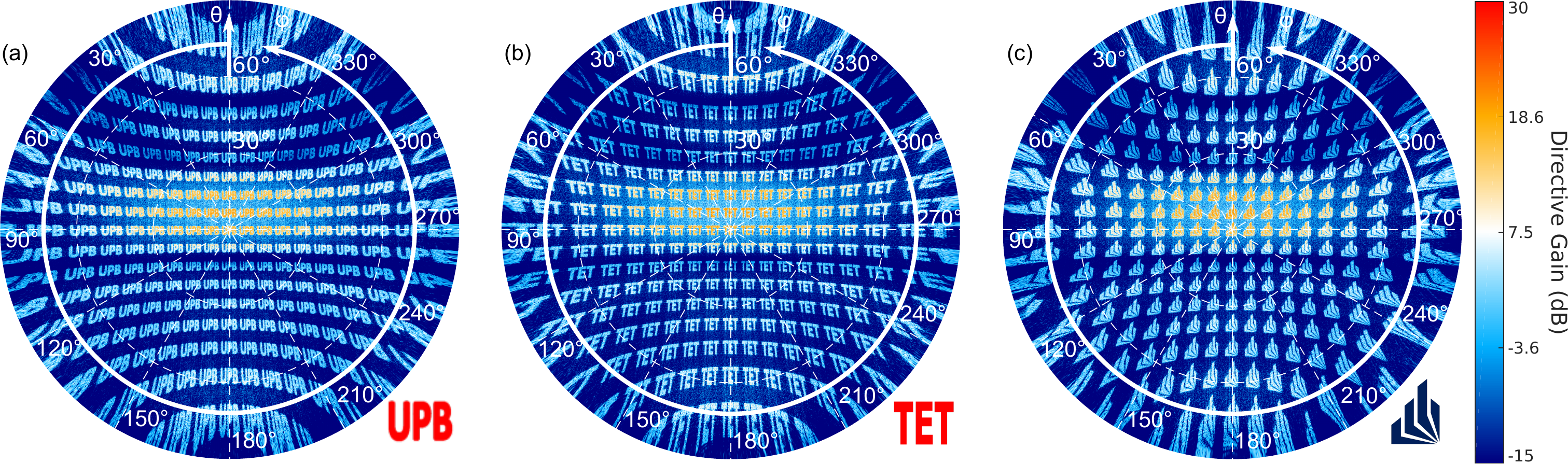}
	\caption{Calculated far-field radiation patterns for a $64 \times 64$ phased array configuration with the optimized antenna to generate (a) the initials ``UPB", b) the initials ``TET", and c) the logo of Paderborn University. The original images used in the process are shown next to the respective far-field patterns.}
	\label{fig:figure3}
\end{figure*}

In the next step, we employ our optimized antenna in a 2D-phased array configuration for which we consider a \SI[mode=text]{9}{\micro\metre}$\times$\SI[mode=text]{9}{\micro\metre} unit cell. As demonstrated in Ref.~\cite{sun2013large1}, the unit cell accommodates a directional coupler and phase shifter for their $64 \times 64$ array configuration. The optimized antenna already has the size that can perfectly fit into this unit cell and also the benefit of much higher upward radiation efficiency compared to the 51\% from the radiator described in the reference. The field for such an array can be defined as 
\begin{equation}
    \mathbf{E}_\mathrm{array}(\theta,\varphi)=\mathbf{E}_\mathrm{antenna}(\theta,\varphi) \, \mathrm{AF}(\theta,\varphi),
\end{equation}
where ${\mathbf{E}_\mathrm{array}(\theta,\varphi)}$ is the far-field of the OPA, ${\mathbf{E}_\mathrm{antenna}(\theta,\varphi)}$ is the far-field of a single antenna, and $\mathrm{AF}(\theta,\varphi)$ is the scalar function representing the array factor. Such arrays can potentially be used in imaging applications, where complex far-field radiation patterns need to be constructed. Fig.~\ref{fig:figure3} illustrates the results of pattern synthesis accomplished by utilizing the optimized antenna in a $64 \times 64$ array. The desired near-field phase distribution for the elements of the uniformly excited phased array is derived using the Gerchberg-Saxton algorithm \cite{fienup1978reconstruction}. Three different images are used for this purpose, namely, the initials of the Paderborn University ``UPB", our department Theoretical Electrical Engineering ``TET", and the Paderborn University logo. The images used for the pattern generation are shown next to their respective far-fields.  Besides, Fig.~\ref{fig:figure3} presents 16 grating lobes in the far-field radiation patterns for each direction due to the large inter-element spacing of \SI[mode=text]{9}{\micro\metre} ($\sim$5.8$\lambda$). This number of interference orders $m$ can be estimated by
\begin{equation}\label{eq3}
    \mid m\lambda_n/d \mid <2,
\end{equation}
where $m$ is the largest value that satisfies Eq.~\ref{eq3}, $\lambda_n$ is the medium wavelength and $d$ is the size of the unit cell. The far-field patterns reveal a brighter region along the horizontal direction, which can be attributed to the large half power beamwidth (HPBW) of $56^\circ$ from each antenna. This large beamwidth is particularly desirable for array configurations that demonstrate beam steering in a range only limited by the FOV of the radiating element \cite{khachaturian2022discretization}. 

\begin{figure*}[t!]
	\centering
	\includegraphics[width=17cm]{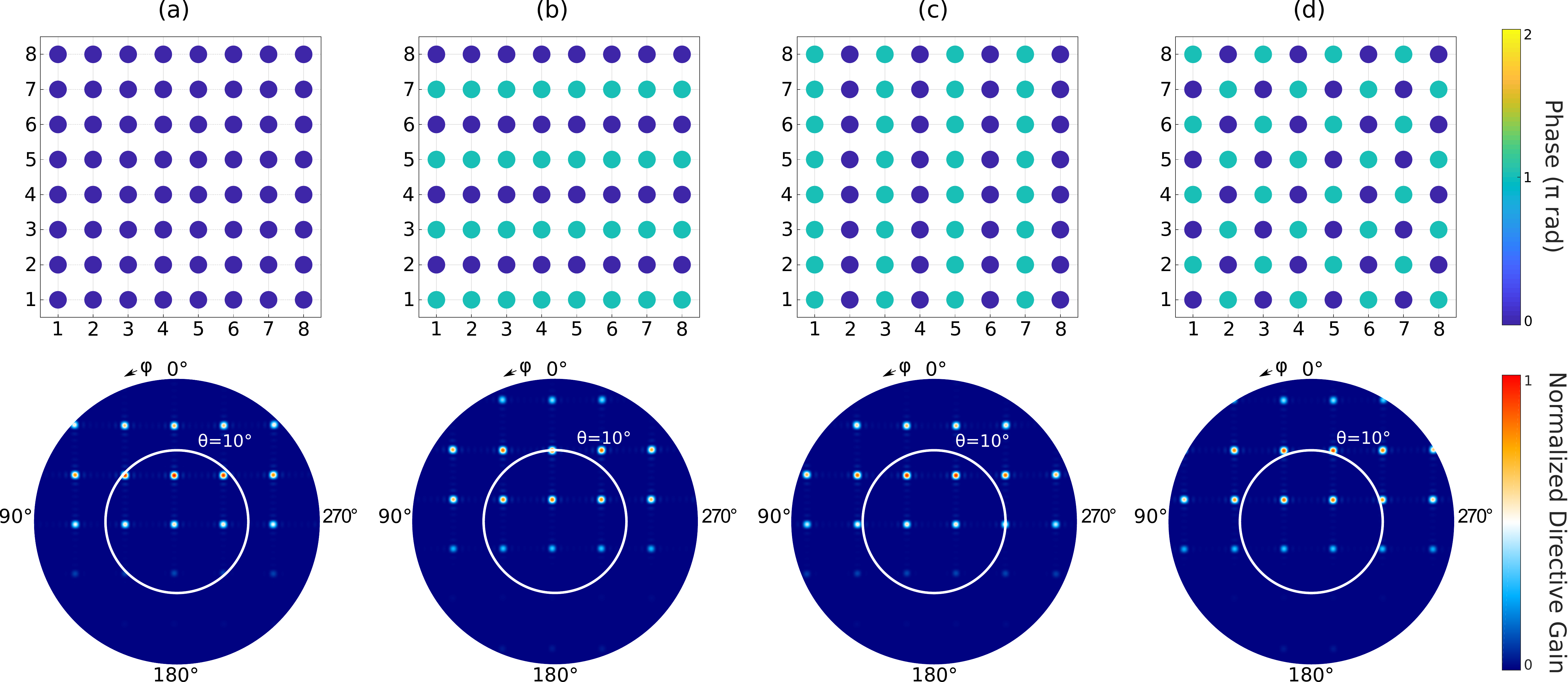}
	\caption{Demonstration of beam steering and beamforming with an $8 \times 8$ array configuration of the optimized antenna. (a) Phase distribution and simulated linear far-field radiation pattern of a uniform phased array. (b-d) Phase distribution and the simulated linear far-field radiation pattern for the main-lobe to be (b) shifted vertically, (c) shifted horizontally, and (d) shifted diagonally.}
	\label{fig:figure4}
\end{figure*}

The large unit cell size of \SI[mode=text]{9}{\micro\metre} also constrains the FOV of the array, which in turn limits the angular range for beam steering. In addition, we illustrate the possibility of steering the beam for which we use an 8$\times$8 OPA that provides better visualization of the steering effect, as seen in Fig.~\ref{fig:figure4}. The far-field patterns are limited to an angular range of $\theta=20^\circ$ to highlight the shifting positions of the main beam. Fig.~\ref{fig:figure4}a demonstrates the phase distribution and far-field pattern for the array when no phase input is applied to the OPA. This serves as the reference for observing the beam steering effect. Alternating the phase with zero and $\pi$ along the rows or columns, as shown in Fig.~\ref{fig:figure4}b and c, steers the beam along the vertical or horizontal directions, respectively. Similarly, alternating the phase inputs along the rows and columns with zero and $\pi$ results in the beam being steered diagonally, as shown in Fig.~\ref{fig:figure4}d.

\begin{table}[h!]
\centering
\caption{Comparison of characteristics of different antennas.}
\begin{tabular}{lllll}
\hline
Design & Footprint & $\eta_{up}$ & $\theta$ & $\sim \text{S}_{11}$\\
\hline
Ref. \cite{sun2013large1}        & \SI[mode=text]{3.5}{\micro\metre}$\times$\SI[mode=text]{2.8}{\micro\metre}   & 51\%  & $15^\circ$ & -13 dB\\
Ref. \cite{fatemi2019nonuniform} & \SI[mode=text]{5}{\micro\metre}$\times$\SI[mode=text]{2}{\micro\metre}       & 51\%  & $7.4^\circ$ & N/A\\
Ref. \cite{liu2022circular}      & \SI[mode=text]{5.5}{\micro\metre}$\times$\SI[mode=text]{2.5}{\micro\metre}   & 71\%  & $6^\circ$& -20 dB\\
Ref. \cite{fatemi2020breaking}   & \SI[mode=text]{5.1}{\micro\metre}$\times$\SI[mode=text]{2}{\micro\metre}     & 35\%  & $9^\circ$ & N/A\\
This work                        & \SI[mode=text]{3.5}{\micro\metre}$\times$\SI[mode=text]{2}{\micro\metre}     & 80\%  & $8^\circ$ & -26 dB\\
\hline
\end{tabular}
  \label{tab:review}
\end{table}

 Finally, Table~\ref{tab:review} compares our results with other antennas used specifically in the context of optical phased array configurations. Our proposed antenna has the smallest footprint and highest efficiency, with a relatively low angle of emission. We accomplish an overall size reduction of 28\% and 49\% in comparison to the footprint reported in Refs.~\cite{sun2013large1} and \cite{liu2022circular}, respectively. Also, an efficiency improvement of 9\% and 45\% is achieved when compared to the radiators from Refs.~\cite{liu2022circular} and \cite{fatemi2020breaking}, respectively. An additional useful metric for comparison is the power reflected back to the waveguide segment, which can be measured using the $S_{11}$ parameter. At \SI[mode=text]{1.55}{\micro\metre}, Refs.~\cite{liu2022circular} and \cite{sun2013large1} report $S_{11}$ values of approximately -20\,dB (1\%) and -13\,dB (5\%). Our proposed structure possesses an $S_{11}$ of -26\,dB (0.25\%), which is a quarter of the acceptable reflected power for many phased array systems \cite{fatemi2020breaking}. Therefore, we envision that any phased array system can operate undisturbed with such a radiating element.

In conclusion, we present the design and optimization of a compact horn-shaped blazed grating antenna that utilizes a heterogeneous grating configuration consisting of a U-shaped grating and L-shaped gratings. The use of a FEM solver in conjunction with an optimization routine reveals a structure with 80\% upward radiation efficiency, an appreciable $18^\circ\times 56^\circ$ full width at half maximum, and negligible reflected power. The proposed antenna is suitable for standard OPAs utilized for pattern synthesis and beamforming, including architectures that introduce the prospect of increasing the grating-lobe-free beam steering range using a large HPBW. Overall, the given antenna design opens up the possibility of fabricating highly efficient phased array systems with desirable radiation characteristics. To our knowledge, the current work presents a radiating element with the highest upward radiation efficiency with the smallest footprint applicable for direct use in 2D-OPAs.

\section*{Acknowledgments} The work was funded by the Ministry of Culture and Science of the state of North Rhine-Westphalia (PhoQC) and Deutsche Forschungsgemeinschaft via TRR142 (C05 \& B06). The authors acknowledge the computing time support provided by the Paderborn Center for Parallel Computing (PC$^2$).

\section*{Data Availability} Data underlying the results presented in this paper are publicly available. (https://doi.org/10.5281/zenodo.7966024)
    
\bibliography{main}
\end{document}